\documentclass[aps,prl, twocolumn,groupedaddress]{revtex4-2}
\usepackage[mathscr]{eucal}
\usepackage{graphicx}
\newcommand{\up}{\uparrow}
\newcommand{\dw}{\downarrow}

\newcommand{\e}{\mathscr{E}}
\newcommand{\h}{\mathscr{H}}
\newcommand{\F}{\mathscr{F}}
\newcommand{\B}{\mathscr{B}}

\begin{document}


\title{Spin Coulomb drag by non-equilibrium magnetic textures}


\author{Igor Lyapilin}
\affiliation{Institute of Metal Physics, UD RAS, Ekaterinburg, 620137, Russia\\
 Ural Federal University, Ekaterinburg}
 %
\date{\today}

\begin{abstract}
Interaction between local magnetization and conduction electrons is
responsible for a variety of phenomena in magnetic materials.  We have shown
that the spin-dependent motive force induced by magnetization dynamics in a
conducting ferromagnet lead to the spin Coulomb drag effect. The spin
Coulomb drag an intrinsic friction mechanism which operates whenever the
average velocities of up-spin and down-spin electrons differ.
\end{abstract}

\maketitle

Introduction: The spatial-temporal dynamics of magnetization $\vec M (\vec
r,t)$ in a ferromagnetic system is described by the phenomenological
Landau-Lifshitz-Gilbert (LLG) equation \cite{Tserk}. Along with this, there is
also an inverse process when the spatial-temporal change in magnetization
induces additional electromagnetic fields acting on conduction electrons. The
spin-driving force (SDF) caused by these fields was first predicted by Berger
\cite{Berger} and has been recently formulated by generalizing Faraday's law
\cite{Barnes}. These fields are non-conservative and spin-dependent. Explicit
expressions for induced spin electromagnetic fields can be represented as
\cite{Yamane, Ohe, Tatara}:
\begin{eqnarray} \label{p1}
\e^{\up\dw}_i = \pm\frac{\hbar}{2e}\vec m\cdot (\partial_t{\vec m}\times\partial_i\vec m),\nonumber\\
\B^{\up\dw}_i = \mp\frac{\hbar}{2e}\epsilon_{ijk}\vec m\cdot (\partial_j{\vec m}\times\partial_k\vec m),
\end{eqnarray}
Here $\vec m = \vec M/|\vec M|$  is a unit vector directed along the
magnetization $\vec M(\vec r, t)$. Arrows $(\up,\dw)$  characterize the
electron spin directions.

The fields in (\ref{p1}) generate a spin-dependent Lorentz force  $\vec
F_s^{\up\dw} =~e\,(\vec{\e}_s +~\vec v \times \vec \B_s^{\up\dw})$. The
manifestation of a spin-driving force is a universal phenomenon in magnetic
metals and can be understood through the gauge field theory, the equation of
motion, and Berry's spin phase \cite{Berry, Stern}. A spin electric field was
experimentally observed in the case of a moving domain wall \cite{Yang}.

Unlike conventional electromagnetic fields, spin electric and magnetic fields
$\vec E_s,$ and\,\,$\vec \B_s$ dramatically affect the kinetics of conduction
electrons. In the field  $\vec E_s$, electrons with spins $\up$, or $\dw$ drift
in opposite directions; thereby inducing a spin-polarized current,   $\vec j_s =
(n^\up-n^\dw)\,e\, \vec v_s.$\,\,Here $e$, and \,$\vec v_s,$ are the charge of
an electron and the drift velocity of the carriers due to the spin electric field
$\e_s$, respectively. When emerging, the spin magnetic field associated with
the Berry phase produces the spin Hall effect that is detected as an anomalous
Hall effect \cite{Nagaosa}. The Lorentz force caused by the spin magnetic
field deflects free charge carriers in opposite directions, depending on the
orientation of the spin $(\up,\dw)$, thereby inducing a Hall voltage and a spin
current if the electron concentrations $n_{\up}\neq n_{\dw}$.

Here special attention should be drawn to one more characteristic feature of
spin electromagnetic fields. As can be seen, the electric and magnetic
components of the spin fields act as a separator, selecting free charge carriers
in the spin direction $(\up,\dw)$. Consequently, the system of conduction
electrons can be treated as a set of two spin subsystems, each of which is
characterized by its own spin direction. Since the electric field governs the
drift velocity of charge carriers, it is obvious that the drift velocity of each of
the subsystems is spin-dependent. It can be assumed that the spin-dependent
fields create conditions under which the manifestations of effects associated
with the Coulomb interaction of two spin subsystems of charge carriers
become possible.

Electron-electron (e-e) interactions play a critical role in various phenomena,
ranging from high-temperature superconductivity, the fractional quantum Hall
effect, a Wigner crystal, a Mott transition, etc. However, both the magnitude
and the influence of this interaction on the kinetic properties of crystals are
difficult to measure. One of the methods that proved to be effective in
measuring scattering rates directly due to the Coulomb interaction is the
Coulomb drag effect of charge carriers. The essence of the effect is to arise a
response as a generated voltage or electric current in a conductive system
when current passes through another conductive film separated from the first
one by a dielectric layer \cite{Rojo} (see literature there). The effect is
underlain by the interlayer Coulomb interaction of conduction electrons
separated by a dielectric.

The drag effect of charge carriers due to the Coulomb interaction exhibits
itself also in a system of spin-polarized charge carriers. Consider the
conductivity of carriers with different spins within a two-channel model
(conductivity along two channels with different spin orientations $(\up,\dw)$
in each channel). Let the drift velocities $v^\up,\,v^\dw$ of carriers in each of
the spin subsystems be different. If other electron scattering sources lack, the
Coulomb interaction preserves the system's total momentum and redistributes
it between charge carriers in different $(\up)$ or $(\dw)$ channels. In this
case, faster carriers drag slow ones through transferring a part of their
momentum to implement the spin Coulomb drag effect (SCD) \cite{Amico}.
Then, it is obvious that the charging current  $j_e = e (n^\up + n^\dw)\,v_e$
($v_e$ is the  drift velocity of electrons) will be unchanged against the spin
current $j_s= (n^\up v^\up - n^\dw v^\dw)$. When the drift velocities of
carriers in the spin channels are equalized, $v_\up =v_\dw$, the spin current is
$j_s=0$,  if $n^\up=n^\dw$.

It should be underscored that for the spin Coulomb drag effect to be observed,
the spin system of conduction electrons divides into two subsystems not
physically but through only the spin orientation $(\up,\dw)$. Such a
representation of the electron system for the manifestation of the SCD effect
is fundamental. Some of the possible scenarios for such a representation are
considered in \cite{Amico}. Note that the concomitant spin electric and
magnetic fields produced in the case of a spatial-temporal inhomogeneous
magnetic structure naturally lead to spin separation of charge carriers to
provide the condition necessary for SCD to arise as the difference in the drift
velocities of electrons in spin channels.

The quantitative measure of SCD is the transresistivity $\rho_{\up\dw}$; it is
defined as the ratio between the electrochemical potential gradient
$\zeta_\up$ for electrons with spin $\up$ and the electric current $j_\dw,
\,\,\,(j_\dw=0).$  This definition is similar to that of the Coulomb drag effect
in two spatially separated layers. The results of calculating $\rho_{\up\dw}$
for a three-dimensional electron gas in the random-phase approximation are
presented in \cite{Rojo, Amico-2}.

Next, we demonstrate that the separation of charge carriers by their spin state,
associated with the concomitant spin electric and magnetic fields, when taking
into account the Coulomb interaction between charge carriers, leads to the
effect of spin Coulomb drag of charge carriers.

The spin Coulomb drag effect can be calculated applying the Kubo method,
Green's functions, or the Boltzmann kinetic equation. However, it is easier to
understand SCD at the phenomenological level.

So, let us look into an electric current flowing through a non-equilibrium
inhomogeneous magnetic system; the latter's magnetization $\vec M(\vec r,t)$
has a spatial-temporal inhomogeneity. The Hamiltonian of the system at hand
is:
$$\h=\h_e +\h_{eE}+\h_{sd} +\h_{ee}+\h_{ev}+\h_v,$$
where  $\h_e=\h_k+\h_s$ is the Hamiltonian of free conduction electrons,
$\h_{eE}$ is interaction with an external electric field $\vec e_0$. $\h_{sd}$
 is exchange interaction between free and localized electrons. $\h_{ee}$ is
Coulomb interaction between conduction electrons. $\h_{ev}$ describes the
interaction with scattering centers. $\h_v$ is the lattice Hamiltonian.

It is assumed that upon performing the Hamiltonian's canonical
transformation that diagonalizes the exchange interaction, we arrive at the
realization of concomitant spin electric  $\e_s$ and magnetic  $\B_s$ fields.
As a consequence, the division of the system of conduction electrons into two
spin subsystems $(\up,\dw)$ takes place.

To start with, we explore the SCD effect purely phenomenologically. For this,
we resort to the equation of motion $m(d\vec \upsilon/dt) = \sum_i\vec F_i$.
The drift velocities of electrons $\vec \upsilon^\up,\,\vec \upsilon^\dw$ are
formed by forces acting on electrons, induced by electric fields and the
Coulomb interaction between electrons:
$$\vec \upsilon ^{\up\dw} =<\frac{(\vec F_{\vec E^*}+\vec F_c) \tau}{m}>$$
Here $ \vec F_{\vec E*}+\vec F_{c}$ are the forces acting on the electrons
from the electric  $\e^*=\vec \e_0 +\e_s$ fields and are also due to the
Coulomb interaction $\vec F_c$. Time $\tau$ characterizes dissipative
processes.

For simplicity, we restrict ourselves to investigation of a spin subsystem of
conduction electrons (for example, with spin $\alpha =\up,\,\dw$). Let
$\upsilon^\alpha$ be the speed of the mass-center,$n^\alpha$ is the number of
such charge carriers. We write down an equation of motion for the electrons
of the spin subsystem with spin $(\alpha=\up,\dw)$ as follows:
\begin{equation}\label{e1}
\dot{p}^\alpha = -e\,n^\alpha \vec\e^*+\F_c^{\alpha,-\alpha} +\dot{p}^\alpha_{ev},
\end{equation}
where  $\vec \e^*=\vec \e_0+\vec \e^\alpha$. $\vec \e_0,\,\,\vec\e_\alpha$ are
the external and spin electric fields. The first summand on the right side of the
equation involves the interaction of electrons with electric fields. The second
summand is responsible for the Coulomb interaction between electrons with
different spin orientations (different spin subsystems). At the same time, it is
obviously that $\F_c^{\up\dw} = -\F_c^{\dw\up}$. Preserving the total
momentum of the electronic system, the Coulomb interaction of electrons
redistributes it between the spin $(\up,\,\dw)$ subsystems of electrons (see
figure). Then $|n^\up \vec v_c| = |n^\dw \vec v_c|$, where $\vec v_c$ is the
momentum lost (acquired) in a single electron-scattering act. The last term
$\dot{p}^\up_{ev}= (i\hbar)^{-1}[p^\up,\,\h_{ev}]$ as relaxational one can
be written in the relaxation time approximation.

Before proceeding to the microscopic description of the SCD
effect, consider it phenomenologically. For this, we write down
expressions for the total charge $j_e$ and spin $j_s$ currents,
believing that each of the forces acting on the conduction
electrons contributes to forming the electron drift velocity.
Besides, the momentum is assumed to be transferred from the
subsystem $\up$ to the subsystem $\dw$ due to the Coulomb
interaction. An expression for the charge current is given as
$$j_e = e [n^\up\,(\vec \upsilon_0 + \vec \upsilon_s - \vec \upsilon_c)
 + n^\dw (\vec \upsilon_0 - \vec \upsilon_s +\vec \upsilon_c)],$$
where $\vec \upsilon_0,\vec \upsilon_s,\vec \upsilon_c$ are the
components of the drift velocity of electrons, due to the action
of the electric fields $\e_0,\,\e_s$ and Coulomb interaction

If the charge current is spin-polarized ($\,n^\up\neq
n^\dw,\,\,n=n^\up+n^\dw$), then
$$\vec j_e = e [n\,\vec \upsilon_0 +(n^\up - n^\dw)(\vec \upsilon_s-
\vec \upsilon_c)] = e n \vec \upsilon_0,$$ e.g. the drag effect
contributes to the spin-polarized current until $ \vec
\upsilon_s \neq\vec \upsilon_c$. When $ \vec \upsilon_s =\vec
\upsilon_c$, the contribution to the spin-polarized current
induced by the spin drag effect vanishes and $ \vec j_e = e
[(n^\up+n^\dw)]\,\vec \upsilon_0.$

Consider an expression for the spin current. We have
$$\vec j_s = e [n^\up\,(\vec \upsilon_0 +\vec \upsilon_s -\vec \upsilon_c)
 - n^\dw (\vec \upsilon_0 -\vec \upsilon_s +\vec \upsilon_c)].$$
or
$$\vec j_s = e [(n^\up-n^\dw)\,\vec \upsilon_0 +n (\vec \upsilon_s - \vec
\upsilon_c).$$ As can be seen, at $n^\up=n^\dw$, the spin
current is nonzero as long as there is a possibility of pumping
the momentum between spin subsystems and $\vec j_s=0$ when $\vec
\upsilon_s=\vec \upsilon_c$. If $n^\up\neq n^\dw$, the SCD
effect in the spin current manifests itself until pumping the
electron momentum between the spin subsystems is possible to
happen. When $\vec \upsilon_s=\vec \upsilon_c$, the effect of
the SCD turns to zero and $ \vec j_s = e [(n^\up-n^\dw)\,\vec
\upsilon_0.$

Using the mass-center approximation, the expression that
determines the Coulomb part of the force $\vec F_c$ can be
written as \cite{Amico}
\begin{equation}\label{a1}
\vec F_c \equiv \vec F^{\up\dw} =-d\, m\,
n^{\up}\,(n^{\dw}/n)\,(\vec \upsilon^\up-\vec \upsilon^\dw),
\end{equation}
where $n=n^\up+ n^\dw$. The coefficient $d$ controls the spin
drag. An explicit expression for the coefficient $d$ comes from
the expression for the resistance matrix $\vec \e^\up
=\sum_\dw\,\rho^{\up\dw}\,\vec j^\dw$. Then
\begin{equation}\label{a2}
d = \frac{ne^2}{m}\rho^{\up\dw}.
\end{equation}
In the Born interaction approximation, the expression for
transresistivity $\rho^{\up\dw}$ can be deduced by various
methods, for example, such as the formalism of the Green's
function method, the method of projection operators \cite{Lyap},
or within the Boltzmann kinetic equation \cite{Amico-3}. The
microscopic theory of calculating the quantity $\rho^{\up\dw}$
is similar to calculating the conventional Coulomb resistance
between parallel two-dimensional layers of an electron or hole
gas \cite{Rojo}. However, it differs in some important aspects.
Firstly, charge carriers with opposite spins interact with the
same set of scattering centers. Secondly, the energy spectrum of
charge carriers becomes spin-dependent as a result of the
canonical transformation of the original Hamiltonian.

In keeping with \cite{Jauho, Lyap}, we write down a general
expression for $\rho^{\up\dw}$:
\begin{eqnarray}\label{r1}
 \rho^{\up\dw} = \frac{\pi\hbar^2}{4\,n^\up n^\dw\, m^2}\,\sum
\limits_{\vec p',\vec p,\vec
q}\,\int\limits_{-\infty}^\infty\,d\omega\,q^2\,|V(q)|^2\,f_1'(\varepsilon_{\vec
p,\up}) \nonumber\\\times (\,f_2(\varepsilon_{\vec p\,',\dw})
-f_2(\varepsilon_{\vec p\,'+\vec q,\,\dw})\,)
\left\{\frac{f_1(\varepsilon_{\vec p-\vec q,\up})}{e^\Delta-1} -
\frac{1-f_1(\varepsilon_{
\vec p-\vec q,\up})}{1-e^{-\Delta}}]\right\}\nonumber\\
\delta(\varepsilon_{\vec p-\vec q,\up}-\varepsilon_{\vec
p,\up}+\hbar\omega)\, \delta(\varepsilon_{\vec p\,'+\vec
q,\dw}-\varepsilon_{\vec p\,',\dw}-\hbar\omega).\qquad\quad
\end{eqnarray}
Here $\Delta=\hbar\omega/k_BT$, $k_B$ is the Boltzmann constant,
$V_q=4\pi^2/q^2\epsilon$ is Fourier component of the interaction
potential, $\epsilon$ is the dielectric constant of material,
$f_i(\varepsilon_{\vec p,\,\sigma}),\,(i=1,2) $ are the
Fermi-Dirac distribution functions for conduction electrons.

Going over from the summation over vectors $\vec p,\,\vec p'$ to
integration and evaluating the integrals, we arrive at an
expression for the degenerate statistics of conduction electrons
and $\hbar\omega \ll k_BT$:
\begin{eqnarray}\label{r1}
\rho^{\up\dw} = \frac{\beta}{6\pi\,n^\up n^\dw\,e^2\,V}\, \,\sum
\limits_{\vec q}\,q^2
\int\limits_{0}^\infty\,d\omega\,\times\nonumber\\|V(q)|^2\,
\frac{\chi^{''}_{0\up}(q,\omega)\,\chi^{''}_{0\dw}(q,-\omega)}
{|\epsilon(q,\omega)|^2\sinh^2(\beta\omega/2)},
\end{eqnarray}
where $\beta=1/k_BT$. $V$ is the volume of the system.
$\chi_{0\alpha}(q,\omega)$ is the noninteracting spin resolved
density-density response function, and
$\varepsilon(q,\omega)=1-V_q\chi_{0\up}(q,\omega)-
V_q\chi_{0\dw}(q\omega)$ is the random-phase approximation
 dielectric function.

Further calculations of the transresistivity $ \rho^{\up\dw} $,
as well as its various approximations, accounting for
dissipation processes and possible methods for detecting the
spin Coulomb drag effect face no difficulties (see, for example,
\cite{Narozhny, Amico-2}).

Finally, we note that the presence of a spin-dependent magnetic
field $\B^{\up\dw}$ acting differently on charged particles with
spin $(\up,\,\dw)$ (shifts them in opposite directions), also
affects the formation of drift velocities of charge carriers,
and, consequently, the charge and spin currents, depending on
the direction of their spin. Thus, the spin-dependent magnetic
field will also manifest itself in the spin Coulomb drag effect.

The research was carried out within the state assignment of Ministry of
Science and Higher Education of the Russian Federation (theme "Spin" No.
122021000036-3).

\end{document}